\documentclass{emulateapj}
\newcommand{\expnt}[2]{\ensuremath{#1 \times 10^{#2}}}   
\newcommand{\psr}{PSR~J1816+4510}

\newcommand{\til}{\raisebox{-0.7ex}{\ensuremath{\tilde{\;}}}}

\newcommand{\kms}{\ensuremath{{\rm km}\,{\rm s}^{-1}}}
\newcommand{\mc}{\multicolumn}
\newcommand{\logg}{\ensuremath{\log_{10}(g)}}
\newcommand{\Teff}{\ensuremath{T_{\rm eff}}}

\begin{document}

\slugcomment{Accepted for publication in the Astrophysical Journal}
\title{A Metal-Rich Low-Gravity Companion to a Massive Millisecond Pulsar}

\author{D.~L.~Kaplan\altaffilmark{1}, V.~B.~Bhalerao\altaffilmark{2,3}, M.~H.~van~Kerkwijk\altaffilmark{4},
  D.~Koester\altaffilmark{5}, S.~R.~Kulkarni\altaffilmark{3}, \&\ K.~Stovall\altaffilmark{6}}

\altaffiltext{1}{Physics Department, University of
  Wisconsin-Milwaukee, Milwaukee WI 53211, USA; kaplan@uwm.edu}
\altaffiltext{2}{Inter-University Centre for Astronomy and Astrophysics, Post Bag 4, Ganeshkhind, Pune 411 007, India}
\altaffiltext{3}{Cahill Center for Astrophysics,
California Institute of Technology, Pasadena, CA 91125, USA}
\altaffiltext{4}{Department of Astronomy and Astrophysics, University
  of Toronto, 60 St.\ George Street, Toronto, ON M5S 3H8, Canada;
mhvk@astro.utoronto.ca}
\altaffiltext{5}{Institut f\"ur Theoretische Physik und Astrophysik, University of Kiel, 24098 Kiel, Germany}
\altaffiltext{6}{Department of Physics and Astronomy, University of
  Texas at San Antonio, San Antonio, TX 78249, USA}

\begin{abstract}
Most millisecond pulsars with low-mass companions are in systems with
either helium-core white dwarfs or non-degenerate (``black widow'' or
``redback'') stars.  A candidate counterpart to \psr\ was identified
by \citet{ksr+12} whose properties were suggestive of both types of
companions although identical to neither.  We have assembled optical
spectroscopy of the candidate companion and confirm that it is part of
the binary system with a radial velocity amplitude of $343\pm7\,\kms$,
implying a high pulsar mass, $M_{\rm
  psr}\sin^3i=1.84\pm0.11\,M_\odot$, and a companion mass
$M_c\sin^3i=0.193\pm0.012\,M_\odot$, where $i$ is the inclination of
the orbit.  The companion appears similar to proto-white dwarfs/sdB
stars, with a gravity $\logg=4.9\pm0.3$, and effective temperature
$16,000\pm500\,$K.  The strongest lines in the spectrum are from
hydrogen, but numerous lines from helium, calcium, silicon, and
magnesium are present as well, with implied abundances of roughly ten
times solar (relative to hydrogen).  As such, while from the spectrum
the companion to \psr\ is superficially most similar to a low-mass
white dwarf, it has much lower gravity, is substantially larger, and
shows substantial metals.  Furthermore, it is able to produce ionized
gas eclipses, which had previously been seen only for low-mass,
non-degenerate companions in redback or black widow systems.  We
discuss the companion in relation to other sources, but find we
understand neither its nature nor its origins.  Thus, the system is
interesting for understanding unusual stellar products of binary
evolution, as well as, independent of its nature, for determining
neutron-star masses.
\end{abstract}


\section{Introduction}
The number and variety of millisecond pulsars (MSPs) has expanded
dramatically in recent years, due to the combination of significant
radio surveys \citep[e.g.,][]{bbb+11,kjb+12,blr+12,lbr+12} and the
advent of the \textit{Fermi} satellite (see \citealt{rap+12} for a
recent review).  These millisecond pulsars offer unprecedented
laboratories to study particle acceleration \citep{aaa+09}, binary
evolution \citep*{bdvh12}, and the masses of both white dwarfs
\citep{cbp+12} and neutron stars \citep{dpr+10,rfs+12}.

Quite a number of discoveries have in fact been the result of the
confluence of these trends, picking out unidentified $\gamma$-ray
sources as priority targets for ongoing radio surveys.  This technique
resulted in the discovery of the eclipsing MSP \object[PSR
  J1816+4510]{\psr} \citep{ksr+12} as part of the ongoing Green Bank
North Celestial Cap (GBNCC) survey (Stovall et~al.\ 2013, in
prep).   This source has a spin-period of 3.2\,ms, an orbital period
of 8.66\,hr, and shows both $\gamma$-ray pulsations and radio eclipses
for $\approx 10$\% of the orbit.  

While the radio and $\gamma$-ray properties resembled a black widow or
redback system \citep{roberts11}, with a non-degenerate companion
being ablated by the pulsar's wind, the blue colors and lack of
variability suggested instead that the companion to \psr\ was more
similar to a helium-core white dwarf \citep{vkbjj05}, even though the
inferred radius seemed too large and the radio eclipses were
unprecedented among those systems.

Here we present optical spectroscopy of the companion to
\psr\ (Section~\ref{sec:spec}), taken with the Keck~I, Palomar 5m, and
William Herschel telescopes.  We confirm through radial velocity
variations that the source identified by \citet{ksr+12} is in fact the
companion (Section~\ref{sec:rv}), and that, at first blush, it resembles a hot,
low-gravity white dwarf (Section~\ref{sec:spec}).  However, detailed
analysis shows that the spectrum has strong lines of helium and metals
(calcium, silicon, magnesium, and possibly iron).  We discuss the
implications of the spectrum and radial velocity measurements in
Section~\ref{sec:discuss}, along with possible origins for the unique
features of this system and ways they can be exploited to learn more
about neutron stars and their companions.  We
conclude in Section~\ref{sec:conc}.

\begin{figure*}
\plottwo{f1a.ps}{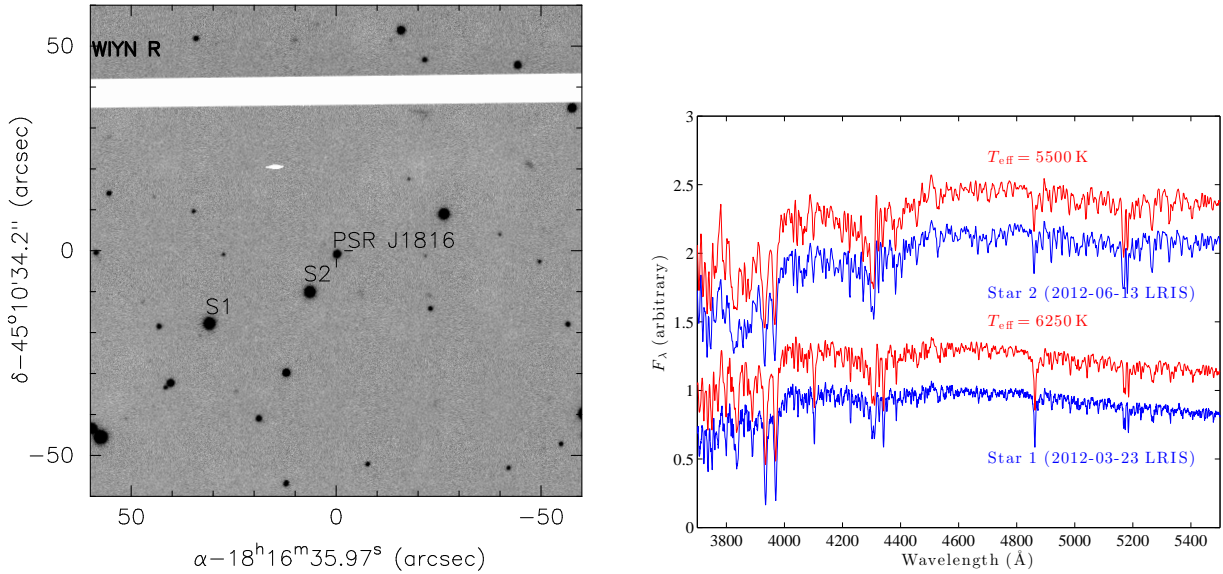}
\caption{Stars used for velocity references.  \textit{Left:} optical
  images of the \psr\ field from \citet{ksr+12}.  The radio position
  is indicated with the tick-marks, and the uncertainties are
  dominated by uncertainties in the absolute astrometry of the optical
  data.  We also label the two reference stars used for our spectra.
  \textit{Right:} spectra of the two reference stars with best-fit
  models.  Star 1 (with the data from the 2012~March Keck
  observations) is at the bottom, along with a 6250\,K model from
  \citet[][offset for clarity]{mscz05}.  Above them is star 2 (data
  from 2012~June Keck observations) and a 5750\,K model. 
}
\label{fig:image}
\end{figure*}

\begin{deluxetable*}{l c c c l c c}
\tablewidth{0pt}
\tablecaption{Summary of Spectroscopic Observations\label{tab:summary}}
\tablehead{
\colhead{Date} & \colhead{UT\tablenotemark{a}} & \colhead{Phase\tablenotemark{b}} & \colhead{Exposure} &
\colhead{Telescope/Instrument\tablenotemark{c}} &
\colhead{Star\tablenotemark{d}} & \colhead{Velocity\tablenotemark{e}} \\
\mc{2}{c}{(TDB BJD)} && \colhead{(sec)} & &  & \colhead{(km/s)}
}
\startdata
2012-02-25 & 06:28 & 0.116 & 900 &  WHT,ISIS+QUCAM,300B & 1 & $-337\pm16$\\
 & 06:41  & 0.141 & 600 & & &$-317\pm20$\\
2012-03-23 & 14:56 & 0.908 & 600 & Keck~I,LRIS-B,600/4000\,\AA & 1 & $-379\pm12$\\
 & 15:07 & 0.930 & 600 & &  & $-396\pm12$\\
2012-04-20 & 10:39 & 0.999 & 600 & P200,DBSP,600/4000\,\AA & 1 & $-457\pm13$\\
2012-05-19 & 07:29 & 0.989 & 360 & P200,DBSP,600/4000\,\AA & 2 & $-444\pm15$\\
& 09:47 & 0.255 & 360 & &&$-102\pm14$\\
& 10:47 & 0.371 & 360 & &&\phs$118\pm14$\\
2012-05-20 & 08:40 & 0.896 & 360& P200,DBSP,600/4000\,\AA & 2 & $-374\pm14$\\
& 10:03 & 0.057 & 360 & && $-444\pm13$\\
& 11:17 & 0.198 & 360 & && $-200\pm14$\\
2012-05-22 & 09:28 & 0.531 & 240 & Keck~I,LRIS-B,400/3400\,\AA & 2 & \phs$251\pm15$\\
2012-06-13 & 07:31 & 0.266 & 240 & Keck~I,LRIS-B,400/3400\,\AA & 2 & \phn$-57\pm14$\\
 & 07:35 & 0.274 & 240 &  && \phn$-25\pm13$\\
2012-08-18 & 09:34 & 0.381 & 300 & Keck~I,LRIS-B,600/4000\,\AA & 1 & \phs$146\pm13$\\
 & 09:40 & 0.394 & 300  & && \phs$165\pm13$\\
\enddata
\tablenotetext{a}{Time of the middle of the exposure, corrected to the
  Solar System barycenter.}
\tablenotetext{b}{Orbital phase, based on the ephemeris of Stovall et
  al.\ (2013, in prep.).}
\tablenotetext{c}{The telescope, instrument, and grating/grism used.
  In most cases we only discuss the results from the blue sides of the
  spectrographs.  ISIS is the Intermediate dispersion Spectrograph and
  Imaging System (ISIS) on the  4.2-m
  William Herschel Telescope, used with the QUCAM electron-multiplying detectors.
 LRIS is the Low Resolution Imaging Spectrograph
  \citep{occ+95} on
  the 10-m Keck~I telescope.  DBSP is the Double Spectrograph on the
  5-m Palomar 200-inch (Hale) telescope.  
}
\tablenotetext{d}{Reference star for each observation, as labeled in Figure~\ref{fig:image}.}
\tablenotetext{e}{Radial velocity for the pulsar, relative to star 1.
The uncertainties include a $\pm12\,\kms$ systematic uncertainty.}
\end{deluxetable*}

\section{Observations \& Reduction}
\label{sec:obs}
We observed the likely counterpart of \psr\ with a number of
instruments over 2012, as summarized in Table~\ref{tab:summary}.  We
primarily report data from the blue sides of the spectrographs,
although the red sides were analyzed as well.  Flux calibrators were
observed with the same setups (\object[GD 153]{GD~153} for the WHT
data, \object[Feige 34]{Feige~34} for the 2012~April P200 data,
\object[BD+33 2642]{BD+33\degr2642} for the Keck data and the 2012 May
P200 data).  The slit width varied between $0\farcs7$ and $1\farcs5$
depending on the conditions.  For all observations of \psr\ except for
the last two Keck observations we did not have the slit at the
parallactic angle as we instead included both the pulsar and a
reference star on the slit to provide a velocity standard (for the
2012 June Keck observations the reference star was instead observed
right after the pulsar).  We used two different stars, as listed in
Table~\ref{tab:summary} and shown in Figure~\ref{fig:image}.

For all of the observations our reduction followed standard
procedures, although we used \texttt{MIDAS} for the WHT data and
\texttt{IRAF} for the other data.  We subtracted biases and overscans
then flattened the data using dome flats, where the shape of the lamp
was removed with a high-order polynomial.  We then extracted the
spectra of the pulsar and the reference star using optimal weighting, subtracting background as measured on both sides of the object
spectrum.  Wavelength calibration was done using arc spectra, and fluxes
were calibrated using the spectrophotometric standards, correcting for
differences in airmass using extinction curves appropriate for each
site.

\begin{figure}
\plotone{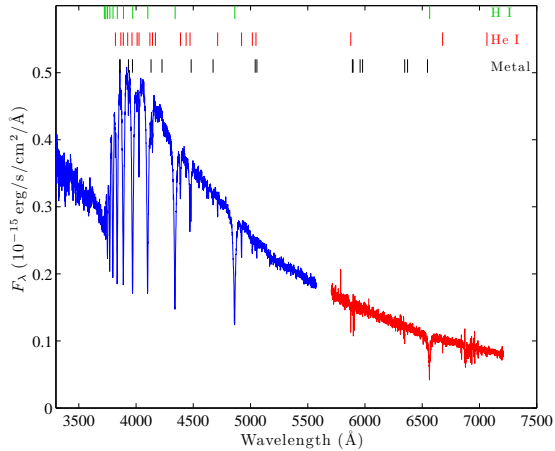}
\caption{Spectrum of the optical counterpart of \psr, based on the
  2012 March Keck observation.  We show the results from both the blue
  and red sides of LRIS.  The  spectrum is reasonably
  consistent with the photometry (\citealt{ksr+12} and Kaplan et
  al.\ 2013, in prep.), given the quality of the flux calibration.
  Along the top are ticks for the different absorption lines seen in
  the spectrum.  Topmost (green) are Balmer lines, as expected from a
  DA white dwarf.  Next (red) are lines from \ion{He}{1}.  Finally
  (black) are low-ionization metal lines (mostly \ion{Ca}{1},
  \ion{Ca}{2}, \ion{Mg}{2}, \ion{Si}{2}).  These are shown in more
  detail in Figure~\ref{fig:normspec}.}
\label{fig:spec}
\end{figure}

\section{Analysis}
\subsection{The Spectrum of \psr}
\label{sec:spec}
\begin{figure*}
\epsscale{0.7}
\plotone{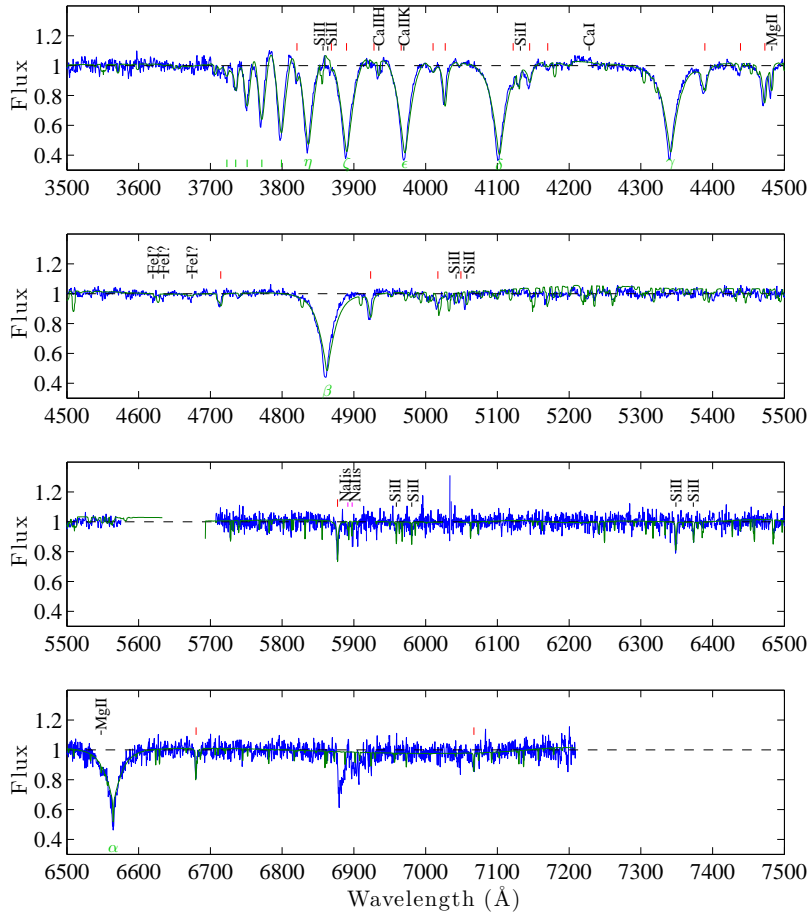}
\caption{Normalized spectrum of the optical counterpart of \psr, based on the
  2012 March Keck observation.  We show the results from both the blue
and red sides of LRIS.  The broad Balmer lines are indicated with
green ticks at the bottom.  \ion{He}{1} lines have red ticks at the
top, while metal lines have black ticks and are labeled by species.
We overplot (green) our best-fit model spectrum.  The region between
5200\,\AA\ and 5500\,\AA\ where the model disagrees slightly with our
data is increasingly affected by uncertain calibration due to the dichroic.}
\label{fig:normspec}
\end{figure*}

To start, we focus on the highest-quality spectra which are the 2012
March Keck data.  We show a set of blue and red-side spectra in 
Figure~\ref{fig:spec}.  One immediately notices the broad Balmer
lines, consistent with our tentative identification of the optical
counterpart as a low-mass white dwarf \citep{ksr+12}.  However, it is
also clear that Balmer lines are not the only ones present.  We also
see strong but narrower lines from \ion{He}{1}, with at least 20 lines
present between 3000\,\AA\ and 7000\,\AA, as well as narrow lines from
\ion{Ca}{2}, \ion{Si}{2}, and \ion{Mg}{2}, and possibly \ion{Fe}{1}
(see Figure~\ref{fig:normspec} for a more detailed view).

\begin{figure}
\plotone{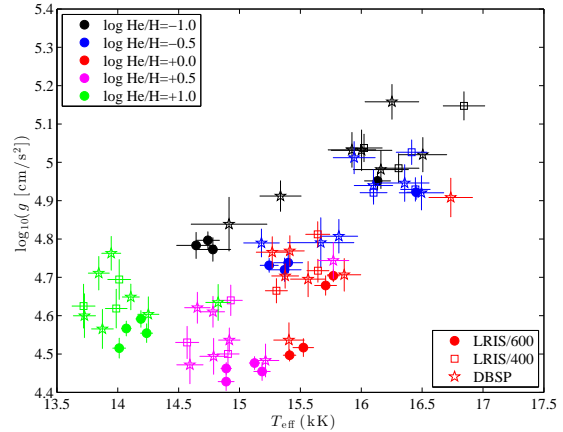}
\caption{Results of model fitting to the companion to \psr.  For each
  value of the helium number density (indicated by the colors, and
  labeled by the log of the number density relative to hydrogen), we
  show the best-fit values of effective temperature and logarithm of
  the surface gravity, determined for all LRIS 600\,l\,mm$^{-1}$
  observations (filled circles), LRIS 400\,l\,mm$^{-1}$
  observations (open squares), and DBSP observations (stars).
  Depending on the exact model there is considerable scatter and/or
  covariance among the best-fit values.  Overall, the model with
  $\log_{10}({\rm He})=0.0$ had the lowest $\chi^2$, followed by
  $\log_{10}({\rm He})=+0.5$, then $\log_{10}({\rm He})=-0.5$.
}
\label{fig:wdfit}
\end{figure}

\begin{deluxetable}{l c c c}
\tablewidth{0pt}
\tablecaption{Elemental Abundances\label{tab:elements}}
\tablehead{
\colhead{Species} & \colhead{$\log_{10}(N/N_{\rm H})$} & \colhead{$\log_{10}(N/N_{\rm
  H})_{\rm Solar}$} & \colhead{$[X/$H$]$}
}
\startdata
He\dotfill& \phs$0.0\pm0.5$ & $-1.0$ & $1.0\pm0.5$\\
Na\tablenotemark{a}\dotfill & $<-3.0$ & $-5.7$ & $<2.7$\\
Mg\dotfill & $-3.8\pm0.2$ & $-4.4$ & $0.6\pm0.2$\\
Si\dotfill & $-3.8\pm0.2$ & $-4.5$ & $0.7\pm0.2$ \\
Ca\dotfill & $-4.5\pm0.3$ & $-5.6$ & $1.1\pm0.3$\\
Fe\tablenotemark{b}\dotfill & $-2.8\pm0.3$ & $-4.5$ & $1.7\pm0.3$\\
\enddata
\tablecomments{Solar abundances are from \citet{ag89}.}
\tablenotetext{a}{Most of the Na seen in the spectrum was
  interstellar, with significant wavelength shifts compared to the
  other species.}
\tablenotetext{b}{Detection of Fe consisted of numerous lines near the
  noise threshold.  Individual lines were not significant, but the
  ensemble was.}
\end{deluxetable}

We fit the data to white-dwarf model
atmospheres to constrain the  atmospheric parameters
$\logg$ and $\Teff$ as well as the He and metal
abundances.  While we obtain reasonable results, we note that the
validity of our models has not really been tested for gravities
$\logg<5.50$ where our best fits are.  Instead those gravities
are more typical of main-sequence/sdB stars.  With this caveat, and
acknowledging the inhomogeneous data-set, our fitting procedure was as
follows.

Our model grid covered gravities $\logg=$4--6 (in steps of 0.25\,dex)
and temperatures $\Teff=14,000$--20,000\,K (in steps of 250\,K).  We
also included helium, with relative abundances $\log_{10} {\rm
  He/H}=-1$ to +1 by number.  For each spectrum we interpolated the
model spectrum to the observed wavelength grid for a range of trial
velocities, convolving the model with the pixel width and a Gaussian
determined by the seeing, and truncating by the slit; we explored also
convolving by the expected radial velocity change during each
observation (see below) but that did not change the results
noticeably.  We normalized the continuum using some line-free regions
and a 3$^{\rm rd}$ order polynomial fit to each spectrum.  The
$\chi^2(\logg,\Teff,v)$ was found by summing over the
3750--4455\,\AA\ region, excluding metal lines, and done separately
for each value of He/H.

After optimizing for velocity (see below), the fit was covariant among
remaining parameters, with the helium abundance altering not just the
line depths but also the continuum shape.  We find reasonable fits for
$\log_{10}{\rm He/H}=-0.5$ to $+0.5$: going to $\log_{10}{\rm He/H}=-1$ (i.e.,
Solar) increases $\chi^2$ by 8670 compared to $\log_{10}{\rm He/H}=0$,
for 9372 degrees-of-freedom.  For each value of the helium abundance,
the best-fit locus in $(\logg,\Teff)$ space was also covariant, with a
positive correlation between the fitted parameters
(Figure~\ref{fig:wdfit}).  Overall more helium resulted in a lower
effective temperature and surface gravity.  Our best-fit was
$\log_{10}{\rm He/H}=0.0\pm0.5$, $\logg=4.9\pm0.3$, and
$\Teff=16,000\pm500\,$K; we believe these uncertainties to be
conservative.  We see no evidence for variations in effective
temperature with orbital phase, down to a limit of about $\pm500\,$K
(limited by the non-uniformity of the data and the uneven phase
sampling).

\begin{figure*}
\epsscale{0.7}
\plotone{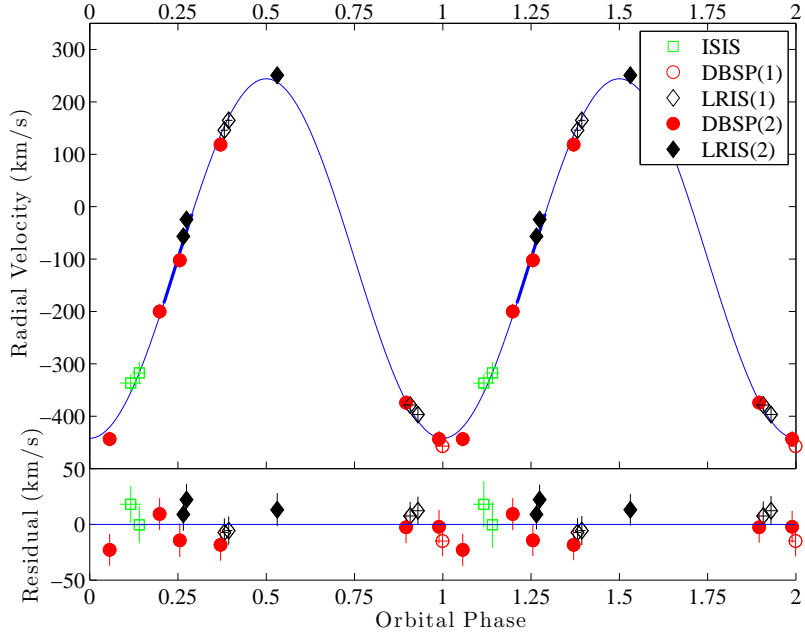}
\caption{Radial velocities of the optical counterpart of \psr\ plotted
  against orbital phase, repeated twice for clarity.  Data from
  WHT/ISIS are squares, P200/DBSP are circles, and Keck~I/LRIS are
  diamonds.  We show data
  using reference star 1 (open symbols) and star 2 (filled
  symbols). The best-fit circular orbit is also shown, with residuals
  in the bottom panel.    The thick portion near phase of 0.25 is the
  approximate phase range where the radio pulsar is eclipsed.}

\label{fig:rv}
\end{figure*}

With the background model determined we examined the metal lines.  We
fit for the abundances of the individual species listed in
Table~\ref{tab:elements} (helium was fit along with the spectrum, as
discussed above).  Other elements were present in our models but they
did not produce significant lines in the 3700--7000\,\AA\ range.
While we could see lines from sodium in the spectrum, they had
different velocities from the rest and therefore are presumably mostly
interstellar, hence we quote only an upper limit.  For iron,
no individual line is significantly detected, but the
large number of lines makes the presence of iron very likely
(Figure~\ref{fig:normspec}).  There may be other species present at
low levels, and we could also determine a number of upper limits, but
given the model uncertainties and our difficulty in interpreting just
the current data we leave the analysis here until the models and data
have improved.  Overall, we find that the metals have abundances
consistent with roughly $10\times$ solar abundance relative to
hydrogen, or roughly solar  relative to helium.

\subsection{The Radial Velocity of \psr}
\label{sec:rv}
We determined radial velocities of the companion to determine its
orbit.  Unfortunately, in the blue part of the spectrum where the
majority of the spectral lines are there are not many sky lines that
can be used as absolute velocity references.  This deficiency is made
more severe by the two-armed nature of the spectrographs used, where
the red- and blue-side wavelength calibrations are separate.  Instead,
we oriented the slit to include both the pulsar's companion and
another reference star.  As discussed above, the observations used two
separate reference stars (Table~\ref{tab:summary}).

Using the models of \citet{mscz05}, we fit for the spectral type of our
reference stars over the wavelength range 3700\,\AA--5400\,\AA,
allowing for an overall  scaling to account for errors in flux
calibration, slit losses, etc.  Reference star \#1 (S1) appears to be
of spectral type mid- to late-F (effective temperature 6250\,K), while
reference star \#2 (S2) is a bit cooler, roughly G2 (5750\,K), as seen
in Figure~\ref{fig:image}.  This is reasonably consistent with the
observed colors (from 2MASS [\citealt{2mass}] and other archival data), where S2 is
slightly redder than S1.

To determine the radial velocities of the pulsar's companion and the
reference star, we used appropriate model atmospheres, minimizing the
$\chi^2$ between the model and the data for a range of trial
velocities.  At each velocity we convolved the model as described
above, and allowed for a low-order polynomial difference between the
model and the data.  We restricted the wavelength range to
3700\,\AA--5400\,\AA\ (the maximum wavelength was 4450\,\AA\ for the WHT
data) so as to focus on a well-calibrated part of the data above the
highest-order Balmer lines and Balmer break but below the dichroic
cutoffs.

For the pulsar, the fits were good, with reduced $\chi^2$ typically
near 1 for the WHT and P200 data.  The higher-quality Keck data had
reduced $\chi^2$ of $\approx 4$, so we scaled the velocity
uncertainties such that the reduced $\chi^2$ was 1.  The fits were
also good for the reference stars, although not quite as good as for
the pulsar.  Instead $\chi^2$ was typically about 1.5 for the P200 and
WHT data and up to 6 for the Keck data.  This may be from a slight
mismatch in resolution (although we explored changing our resolution),
but more likely is from having an imperfect  model grid that is coarse and that
may not reflect the true metallicity or rotation.  In any case, the
best-fit velocities did not depend much on the exact model or the
choices of the other parameters.

The use of two reference stars complicates our analysis slightly,
as each star can of course have its own peculiar velocity.  We
therefore obtained two spectra where S1 and S2 were on the
slit together without the pulsar.  These exposures were right before
the observations of the pulsar on 2012~August~18 and used the same
configuration and reduction.  We found a consistent radial velocity
offset of $-54\pm1\,\kms$ for S2 relative to S1.  In what
follows, we apply this offset to determine the velocities of the
pulsar relative to S1.

We show the best-fit radial velocities in Figure~\ref{fig:rv}.  The
systemic velocity is relative to S1 (in no case were the finite
durations of the exposures significant compared to the orbital period,
with corrections of $<1\,\kms$).  We have included a $\pm 12\,\kms$
systematic velocity uncertainty in the data to account for limitations
in our reduction and differences between the instruments.  This will
also account for mis-alignment of the slit intended to cover two
objects (an offset of approximately one half of the slit width
corresponds to a velocity shift of approximately $10\,\kms$).  With
that, we were able to fit a circular orbit using the phase established
by the radio ephemeris (Stovall et al.), which has since been
confirmed in $\gamma$-rays \citep{ksr+12} and is more accurate than an
optical-only ephemeris.  With all of the data we find a radial
velocity amplitude $K_{\rm c}=343\pm7\,\kms$ for the companion, with
$\chi^2=14.5$ for 14 degrees of freedom and a systemic velocity of
$-99\pm8\,\kms$ relative to S1.  If we only use the P200 data
including  S2---a somewhat more homogeneous data-set---we
find a  consistent $K_{\rm c}=339\pm11\,\kms$ ($\chi^2=3.5$ for 4 degrees of freedom).  We can
also reference all of the velocities to the 5577\,\AA\ sky line.  Here
we might worry that the line is at the extreme edge of our spectra
where the wavelength solutions are not as reliable, but the result is
consistent: $K_{\rm c}=347\pm11\,\kms$ and a systemic velocity of $-108\pm12\,\kms$ ($\chi^2=20.9$
for 12 degrees of freedom).

\section{Discussion}
\label{sec:discuss}
Combining the best-fit radial velocity amplitude of the
companion, $K_c=343\pm7\,\kms$,  with the measured projected semi-major
axis of the pulsar $x_{\rm PSR}=0.595402\pm0.000002\,$s, we can infer
the mass-ratio:
\[
q\equiv \frac{M_{\rm PSR}}{M_{\rm c}}=\frac{K_{\rm c}P_{\rm B}}{2\pi x_{\rm
    PSR} c}=9.54\pm0.21.
\]
The minimum masses for the pulsar and companion are then: $M_{\rm
  psr}\sin^3 i=1.84\pm0.11\,M_\odot$ and $M_{\rm c}\sin^3
i=0.193\pm0.012\,M_\odot$, where $i$ is the inclination of the orbit.
Note that, unlike black widow systems that are strongly affected by
heating such that the photocenter and center of mass are not the same
(\citealt*{vkbk11}; \citealt{rfs+12}), the minimal modulation seen for
\psr\ \citep[][Kaplan et al.\ 2013, in prep]{ksr+12}, suggests that we
are measuring the true center-of-mass motion.

Combining this mass determination with our
spectroscopically-determined gravity, we can determine the radius of
the companion and the distance to the system.  In what follows, we
give the values for the minimum mass, equivalent to that for an
edge-on orbit, which is not unreasonable given the eclipses (although
since they are radio eclipses from ionized material, constraints on
the inclination are weak).  The minimum radius is
$0.26\pm0.08\,R_\odot$, implying an average density of 15\,g\,cm$^{-3}$.  To get the distance, we use the bolometric
corrections appropriate for this effective temperature from
\citet[][using the 0.2\,$M_\odot$ models]{tbg11}\footnote{See
  \url{http://www.astro.umontreal.ca/{\til}bergeron/CoolingModels/}.},
with $M_{\rm bol}-R=-1.52\pm0.09\,$mag; while this is for a larger
gravity ($\logg=6.6$) than what we measure, the dependence of the
bolometric correction on gravity is quite small.  We assume an
extinction $A_V=0.22\,$mag, which comes from fitting our photometry to
the same synthetic photometry as we used for the bolometric
corrections\footnote{We note that $A_V=0.22\,$mag is somewhat less
  than we had assumed in \citet{ksr+12}, but it is more consistent
  with the total Galactic extinction inferred from molecular gas.  The
  large values of $A_V$ in \citet{ksr+12} were primarily the result of
  fitting the optical and \textit{GALEX} ultraviolet photometry
  simultaneously.  With better data it appears that the models
  significantly overpredict the ultraviolet photometry, and when we
  fit only the optical data we find values of $A_V$ near 0.2\,mag
  (Kaplan et al.\ 2013, in prep).}.  Overall we find
$d=4.5\pm1.7\,$kpc using the $R$-band photometry from \citet{ksr+12},
where the uncertainty is dominated by the uncertainty on the measured
gravity and temperature ($d\propto \Teff^{0.75}10^{-\logg/2}$), and we get
similar results integrating our model atmosphere directly.  The
distance is almost a factor of two larger than the dispersion-measure
distance discussed in \citet{ksr+12}, suggesting that, as is found for
pulsars out of the Galactic plane generally \citep{roberts11}, the
dispersion-measure distance is an underestimate.  The nominal inferred
radius is correspondingly larger, indicating that this source is even
further from equilibrium than previously thought with a thermal
timescale of $\approx 1\,$Myr (the radii of
$0.2\,M_\odot$ WDs are $\lesssim 0.03\,R_\odot$), although it is still
well within the Roche-lobe radius of $0.56\,R_\odot$.  
The distance also implies a larger $\gamma$-ray luminosity,
$L_\gamma=\expnt{5}{34}\,{\rm erg\,s}^{-1}$, roughly equal to the
total spin-down luminosity, $\dot E=\expnt{5}{34}\,{\rm erg\,s}^{-1}$ (ignoring possible
corrections for secular acceleration and assuming the standard moment
of inertia $I=10^{45}\,{\rm g\,cm}^2$), so $L_\gamma /
\dot E\approx 1.0_{-0.7}^{+0.9}$. Note that the $L_\gamma < \dot E$
requirement is somewhat rough, with observed values of $L_\gamma/\dot
E$ (based on poorly-known dispersion measure distances, and assuming
beaming into $4\pi\,$ster) ranging from
10\% to $>200$\% \citep{rrc+11}; given the uncertainty in our
photometric distance we see no problems.

For the nominal $\dot E$, we expect $\approx \expnt{1.1}{11}\,{\rm
  erg\,s^{-1}\,cm}^{-2}$ of incident flux at the surface of the
companion.  This compares with $\sigma \Teff^4=\expnt{3.7}{12}\,{\rm
  erg\,s^{-1}\,cm}^{-2}$ coming from the companion itself, so we would
expect variations of at most $\pm 100\,$K coming from the day and night sides
of the companion (assuming an albedo of 100\%).  Our spectra cannot yet constrain such a variation,
but this should be possible in the future with high signal-to-noise spectroscopy or photometry.

\begin{figure*}
\epsscale{0.7}
\plotone{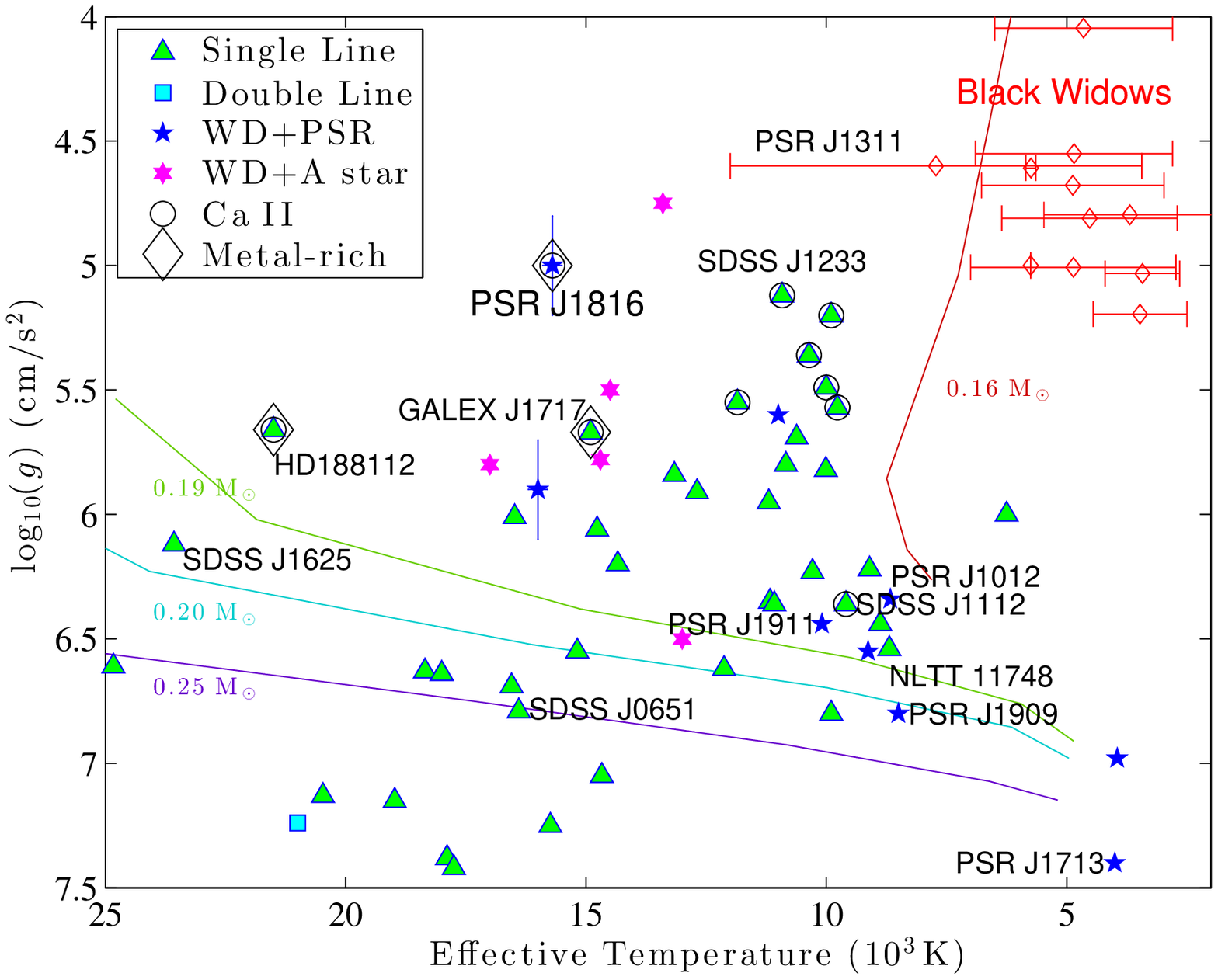}
\caption{Effective temperature vs.\ log(surface gravity) for low-mass
  white dwarfs.  The WD-WD binaries (largely from the ELM surveys of
  \citealt{kbap+12}) are the triangles/squares. Binaries with A stars
  (such as from \citealt{vkrb+10}) are 6-pointed stars, while WD-PSR
  binaries are 5-pointed stars.  All sources with high-quality spectra
  and $\logg\leq 5.6$ have \ion{Ca}{2} in their atmospheres (circles);
  those with $\Teff>15$,000\,K have other species as well (diamonds).
  The evolutionary tracks for $M<0.25\,M_\odot$ are from
  \citet{pach07}, and show only the terminal cooling, ignoring shell
  flashes. \psr\ is labeled, as are other sources of interest: the
  ``proto-WD'' systems GALEX~J171708.5+675712 \citep{vtk+11} and
  HD~188112 \citep{heln03}, pulsars \citep{vkbjj05} and select WD-WD
  binaries \citep{kbap+12}. SDSS~J111215.82+111745.0 is also labeled: it
  is notable for having $\logg>6$ but still showing \ion{Ca}{2} and
  for having detectable pulsations that can be used to help determine
  the depth of the convective zone \citep{hmw+12b}.  We also show some
  black widow/redback systems, as compiled by \citet{bvkr+13}, with the addition of PSR~J1311$-$3430 from \citet{rfs+12}.}
\label{fig:logg}
\end{figure*}

Our very low gravity brings up a question of nomenclature. Initially
we called the companion a white dwarf, and indeed we fit the spectra
with white dwarf model atmospheres.  But this gravity is really more
consistent with a sdB star or a proto-WD, such as \object[GSC
  04421-02656]{GALEX~J171708.5+675712} \citep{vtk+11} or \object[HD
  188112]{HD~188112} \citep{heln03}.  As shown in
Figure~\ref{fig:logg}, \psr\ is even more extreme in the gravity axis
than those sources.  It almost certainly is not a standard sdB star,
as these typically have effective temperatures $>25$,000\,K and masses
of $\approx 0.5\,M_\odot$ (able to sustain core He fusion) compared to
$16$,000\,K and $0.2\,M_\odot$ here\footnote{While the inclination is unknown
so in principle a companion mass of $\approx 0.5\,M_\odot$ is
possible, our constraint on the mass ratio would then require a
neutron star mass of $>5\,M_\odot$}.  It may still be contracting and
cooling from a recent hydrogen shell flash, as discussed in
\citet{ksr+12}.  We note that the approximate mass of $0.2\,M_\odot$
is at the limit where He-core WDs are expected to undergo shell
flashes instead of steady burning \citep{pach07}: if we could
determine a more precise mass and establish whether or not unstable
flashes occurred, it will help our understanding of the evolution
of low-mass WDs.

Another question is the origin of the metals\footnote{If interpreted
  as a white dwarf, the companion would be a DABZ star (hydrogen
  atmosphere with significant helium and metals), which is an almost
  unique combination, especially at this gravity.  The companion is in
  fact more metal-rich than the previous record holder,
  \object[SDSS J073842.56+183509.6]{SDSS~J073842.56+183509.6} \citep{dkf+10}.}.  While planetesimal
accretion has been suggested for other low-gravity WDs
(Figure~\ref{fig:logg}; \citealt{gkf+12}), this seems very unlikely
for \psr, where the relativistic wind of the pulsar and the ionized
wind of the companion will both suppress accretion.  The solar Ca/He
ratio also supports a primordial origin for the metals
(cf.\ \citealt{dkf+10}).  At the other mass extreme, comparison with
sdB stars also shows differences.  For instance, \citet{ehh+03} show
that helium abundance increases with effective temperature, becoming
close to solar only for $\Teff>30$,000\,K.  \citet{geier12} finds no
correlation of metal abundances with \Teff\ or the helium abundance,
although they are generally sub-solar. The ionized wind that likely
causes the radio eclipses may be part of the answer here, as it might
alter diffuse equilibrium (much as inferred for sdB stars winds;
\citealt{ub98}) and hence the onset of shell burning.  It also may
remove enough angular momentum to create ultra-compact binaries such
as \object[PSR J1719-14]{PSR~J1719$-$1438} and its planetary-mass
companion \citep{vhnvj12}.  Alternatively, if a shell flash occurred
recently, this might not only explain the low surface gravity but also
the abundances, as it might have mixed up the interior and the present
low gravity would delay sedimentation compared to an ordinary white
dwarf.

Another interesting comparison is the black widow system \object[PSR
  J1311-3430]{PSR~J1311$-$3430} \citep{rfs+12}.  Its companion has a
lower gravity, a much lower mass, and is likely less degenerate than
the companion of \psr, yet has a comparable maximum temperature and a
spectrum dominated by lines of \ion{He}{1}, with small amounts of
metals (\ion{Ca}{2}, \ion{Mg}{2}) visible and no detectable hydrogen.
Could \psr\ evolve into a system like PSR~J1311$-$3430?  Current
suggestions for the origin of PSR~J1311$-$3430 involve substantial
radiative stripping \citep{bdvh12}, rather than unstable hydrogen
burning, so perhaps not.  Indeed, there are indications for a dense
wind from PSR~J1311$-$3430 that appears driven by irradiation;
irradiation appears less important for \psr, but detailed calculations
are necessary to determine this quantitatively.

The mass of the neutron star is considerably above the canonical value
of $1.4\,M_\odot$.  This is consistent with the accretion necessary to
have spun it up to its current period \citep*{tlk12} and with the
masses of a number of other pulsars with low-mass He white dwarf
companions \citep{vkbjj05,cbp+12}, almost to the level of black-widow
systems \citep{vkbk11,rfs+12}.  While our current measurement is not
as extreme as some \citep{dpr+10}, the brightness of the companion and
its narrow spectral lines mean that we can potentially get precision
of $<1\%$ on $K_c$ and hence on the mass ratio without uncertainty
about the photocenter \citep[cf.][]{vkbk11}. At this point our
uncertainty in the masses will be entirely dominated by uncertainties
in the inclination.  A Shapiro delay measurement is likely impossible
because of the eclipses, so we may have to rely instead on combining
spectroscopy with modeling the photometric light-curve
\citep{rcf+07,rfs+12}; we have begun detailed time-resolved photometry
to search for irradiation or ellipsoidal variations (expected to be
roughly 2\%).  Given the masses we infer, we can expect the system to
merge due to gravitational radiation in about 1\,Gyr \citep{st83}; the
actual merger time might be shorter if energy is dissipated through
tidal interactions, and the bloated nature of the companion could make
this more significant than otherwise inferred \citep{bqaw12}.

A related possibility is that we might be able to determine the
rotation of the companion, which would help understand the history of
mass transfer in the system \citep[e.g.,][]{pkr+12}.  If rotating synchronously
with the orbit, we would expect the equatorial velocity to be $\approx
35\,\kms$.  This high velocity combined with the low surface gravity
and the presence of narrow He lines means that we might separate the pressure
broadening responsible for the Balmer lines from rotation.  Initial
estimates suggest this might be possible, but again it requires
considerably better data than we have right now along with improved
atmospheric modeling.

\section{Conclusions}
\label{sec:conc}
We have described an initial spectroscopic study of the companion to
the millisecond pulsar \psr.  Our findings largely agree with the
initial speculation from \citet{ksr+12}: that the companion is hot and
large, consistent with neither black widow companions nor with
standard white dwarfs.  However, we now find that it is even more
distinct, in showing strong lines of helium and metals in its
spectrum, together with the abnormally low gravity and ionized-gas
eclipses.

We are left with a number of questions.  (1) What is the lifetime of
such an object?  It cannot be too small compared to typical MSP ages,
since otherwise it would be too unlikely to find it among the
$\sim\!100$ known MSP systems. (2) Why are there strong metals and helium?  Are
they related?  Is the cause of metals similar to that of normal
low-mass WDs, or is it unique to this system?  (3) What are the
detailed metal abundances?  Are they all the same relative to solar?  (4) Why is it
eclipsing?  A temperature of 16,000\,K is generally thought to be too
low for a wind.  Is that related to the pulsar (is it ablating)?  Or
is it a more general property of the companion?  Finally, (5) does the
presence of the pulsar significantly affect (through its relativistic
wind) the evolution of the companion?  These questions have important
consequences for our understanding of white dwarf and neutron star
evolution, as well as the study of interacting binary systems.  

\acknowledgements Partially based on observations made with the
William Herschel Telescope operated on the island of La Palma by the
Isaac Newton Group in the Spanish Observatorio del Roque de los
Muchachos of the Instituto de Astrofísica de Canarias.  Some of the
data presented herein were obtained at the W.~M.~Keck Observatory,
which is operated as a scientific partnership among the California
Institute of Technology, the University of California, and NASA; the
Observatory was made possible by the generous financial support of the
W.~M.~Keck Foundation.  The authors wish to recognize and acknowledge
the very significant cultural role and reverence that the summit of
Mauna Kea has always had within the indigenous Hawaiian community. We
are most fortunate to have the opportunity to conduct observations
from this mountain. We thank David Levitan, Assaf Horesh, and Yi Cao
for assistance with observations; research at Caltech was funded by a
grant from the NSF.  We thank an anonymous referee for helpful
suggestions.

\facility{Keck:I (LRIS)}, \facility{Hale (Double Spectrograph)},
\facility{ING:Herschel (ISIS)}



\begin{thebibliography}{}

\bibitem[{Abdo} {et~al.}(2009){Abdo}, {Ackermann}, {Ajello}, {Atwood},  {Axelsson}, {Baldini}, {Ballet}, {Barbiellini}, {et~al.}]{aaa+09}
{Abdo}, A.~A., {et al.} 2009, Science, 325,  848

\bibitem[{Anders} \& {Grevesse}(1989){Anders} \& {Grevesse}]{ag89}
{Anders}, E. \& {Grevesse}, N. 1989, \gca, 53, 197

\bibitem[{Bailes} {et~al.}(2011){Bailes}, {Bates}, {Bhalerao}, {Bhat},  {Burgay}, {Burke-Spolaor}, {D'Amico}, {Johnston}, {Keith}, {Kramer},  {Kulkarni}, {Levin}, {Lyne}, {Milia}, {Possenti}, {Spitler}, {Stappers}, \&  {van Straten}]{bbb+11}
{Bailes}, M., {et al.} 2011, Science, 333,  1717

\bibitem[{Benvenuto} {et~al.}(2012){Benvenuto}, {De Vito}, \&  {Horvath}]{bdvh12}
{Benvenuto}, O.~G., {De Vito}, M.~A., \& {Horvath}, J.~E. 2012, \apjl, 753, L33

\bibitem[{Boyles} {et~al.}(2012){Boyles}, {Lynch}, {Ransom}, {Stairs},  {Lorimer}, {McLaughlin}, {Hessels}, {Kaspi}, {Kondratiev}, {Archibald},  {Berndsen}, {Cardoso}, {Cherry}, {Epstein}, {Karako-Argaman}, {McPhee},  {Pennucci}, {Roberts}, {Stovall}, \& {van Leeuwen}]{blr+12}
{Boyles}, J., {et al.} 2013, \apj, 763, 80

\bibitem[{Breton} {et~al.}(2013){Breton} {et al.}]{bvkr+13}
{Breton}, R.~P., {van Kerkwijk}, M.~H., {Roberts}, M.~S.~E., {et al.}
2013, \apj, submitted

\bibitem[{Burkart} {et~al.}(2012){Burkart}, {Quataert}, {Arras}, \&  {Weinberg}]{bqaw12}
{Burkart}, J., {Quataert}, E., {Arras}, P., \& {Weinberg}, N.~N. 2012, \mnras,  submitted, arXiv:1211.1393

\bibitem[{Corongiu} {et~al.}(2012){Corongiu}, {Burgay}, {Possenti}, {Camilo},  {D'Amico}, {Lyne}, {Manchester}, {Sarkissian}, {Bailes}, {Johnston},  {Kramer}, \& {van Straten}]{cbp+12}
{Corongiu}, A., {et al.} 2012, \apj, 760, 100

\bibitem[{Demorest} {et~al.}(2010){Demorest}, {Pennucci}, {Ransom}, {Roberts},  \& {Hessels}]{dpr+10}
{Demorest}, P.~B., {Pennucci}, T., {Ransom}, S.~M., {Roberts}, M.~S.~E., \&  {Hessels}, J.~W.~T. 2010, \nat, 467, 1081

\bibitem[{Dufour} {et~al.}(2010){Dufour}, {Kilic}, {Fontaine}, {Bergeron},  {Lachapelle}, {Kleinman}, \& {Leggett}]{dkf+10}
{Dufour}, P., {Kilic}, M., {Fontaine}, G., {Bergeron}, P., {Lachapelle}, F.-R.,  {Kleinman}, S.~J., \& {Leggett}, S.~K. 2010, \apj, 719, 803

\bibitem[{Edelmann} {et~al.}(2003){Edelmann}, {Heber}, {Hagen}, {Lemke},  {Dreizler}, {Napiwotzki}, \& {Engels}]{ehh+03}
{Edelmann}, H., {Heber}, U., {Hagen}, H.-J., {Lemke}, M., {Dreizler}, S.,  {Napiwotzki}, R., \& {Engels}, D. 2003, \aap, 400, 939

\bibitem[{G{\"a}nsicke} {et~al.}(2012){G{\"a}nsicke}, {Koester}, {Farihi},  {Girven}, {Parsons}, \& {Breedt}]{gkf+12}
{G{\"a}nsicke}, B.~T., {Koester}, D., {Farihi}, J., {Girven}, J., {Parsons},  S.~G., \& {Breedt}, E. 2012, \mnras, 424, 333

\bibitem[{Geier}(2012){Geier}]{geier12}
{Geier}, S. 2013, \aap, 549, 110

\bibitem[{Heber} {et~al.}(2003){Heber}, {Edelmann}, {Lisker}, \&  {Napiwotzki}]{heln03}
{Heber}, U., {Edelmann}, H., {Lisker}, T., \& {Napiwotzki}, R. 2003, \aap, 411,  L477

\bibitem[{Hermes} {et~al.}(2012){Hermes}, {Montgomery}, {Winget}, {Brown},  {Gianninas}, {Kilic}, {Kenyon}, {Bell}, \& {Harrold}]{hmw+12b}
{Hermes}, J.~J., {et al.} 2012, \apj, in press, arXiv:1211.1022

\bibitem[{Kaplan} {et~al.}(2012){Kaplan}, {Stovall}, {Ransom}, {Roberts},  {Kotulla}, {Archibald}, {Biwer}, {Boyles}, {Dartez}, {Day}, {Ford}, {Garcia},  {Hessels}, {Jenet}, {Karako}, {Kaspi}, {Kondratiev}, {Lorimer}, {Lynch},  {McLaughlin}, {Rohr}, {Siemens}, {Stairs}, \& {van Leeuwen}]{ksr+12}
{Kaplan}, D.~L., {et al.} 2012, \apj, 753, 174

\bibitem[{Keith} {et~al.}(2012){Keith}, {Johnston}, {Bailes}, {Bates}, {Bhat},  {Burgay}, {Burke-Spolaor}, {D'Amico}, {Jameson}, {Kramer}, {Levin}, {Milia},  {Possenti}, {Stappers}, {van Straten}, \& {Parent}]{kjb+12}
{Keith}, M.~J., {et al.} 2012, \mnras, 419, 1752

\bibitem[{Kilic} {et~al.}(2012){Kilic}, {Brown}, {Allende Prieto}, {Kenyon},  {Heinke}, {Ag{\"u}eros}, \& {Kleinman}]{kbap+12}
{Kilic}, M., {Brown}, W.~R., {Allende Prieto}, C., {Kenyon}, S.~J., {Heinke},  C.~O., {Ag{\"u}eros}, M.~A., \& {Kleinman}, S.~J. 2012, \apj, 751, 141

\bibitem[{Lynch} {et~al.}(2012){Lynch}, {Boyles}, {Ransom}, {Stairs},  {Lorimer}, {McLaughlin}, {Hessels}, {Kaspi}, {Kondratiev}, {Archibald},  {Berndsen}, {Cardoso}, {Cherry}, {Epstein}, {Karako-Argaman}, {McPhee},  {Pennucci}, {Roberts}, {Stovall}, \& {van Leeuwen}]{lbr+12}
{Lynch}, R.~S., {et al.} 2013, \apj, 763, 81

\bibitem[{Munari} {et~al.}(2005){Munari}, {Sordo}, {Castelli}, \&  {Zwitter}]{mscz05}
{Munari}, U., {Sordo}, R., {Castelli}, F., \& {Zwitter}, T. 2005, \aap, 442,  1127

\bibitem[{Oke} {et~al.}(1995){Oke}, {Cohen}, {Carr}, {Cromer}, {Dingizian},  {Harris}, {Labrecque}, {Lucinio}, {Schaal}, {Epps}, \& {Miller}]{occ+95}
{Oke}, J.~B., {et al.} 1995, \pasp, 107, 375

\bibitem[{Pablo} {et~al.}(2012){Pablo}, {Kawaler}, {Reed}, {Bloemen},  {Charpinet}, {Hu}, {Telting}, {{\O}stensen}, {Baran}, {Green}, {Hermes},  {Barclay}, {O'Toole}, {Mullally}, {Kurtz}, {Christensen-Dalsgaard},  {Caldwell}, {Christiansen}, \& {Kinemuchi}]{pkr+12}
{Pablo}, H., {et al.} 2012, \mnras, 422, 1343

\bibitem[{Panei} {et~al.}(2007){Panei}, {Althaus}, {Chen}, \& {Han}]{pach07}
{Panei}, J.~A., {Althaus}, L.~G., {Chen}, X., \& {Han}, Z. 2007, \mnras, 382,  779

\bibitem[{Ransom} {et~al.}(2011){Ransom}, {Ray}, {Camilo}, {Roberts}, {{\c  C}elik}, {Wolff}, {Cheung}, {Kerr}, {Pennucci}, {DeCesar}, {Cognard}, {Lyne},  {Stappers}, {Freire}, {Grove}, {Abdo}, {Desvignes}, {Donato}, {Ferrara},  {Gehrels}, {Guillemot}, {Gwon}, {Harding}, {Johnston}, {Keith}, {Kramer},  {Michelson}, {Parent}, {Saz Parkinson}, {Romani}, {Smith}, {Theureau},  {Thompson}, {Weltevrede}, {Wood}, \& {Ziegler}]{rrc+11}
{Ransom}, S.~M., {et al.} 2011, \apjl, 727, L16

\bibitem[{Ray} {et~al.}(2012){Ray}, {Abdo}, {Parent}, {Bhattacharya},  {Bhattacharyya}, {Camilo}, {Cognard}, {Theureau}, {Ferrara}, {Harding},  {Thompson}, {Freire}, {Guillemot}, {Gupta}, {Roy}, {Hessels}, {Johnston},  {Keith}, {Shannon}, {Kerr}, {Michelson}, {Romani}, {Kramer}, {McLaughlin},  {Ransom}, {Roberts}, {Saz Parkinson}, {Ziegler}, {Smith}, {Stappers},  {Weltevrede}, \& {Wood}]{rap+12}
{Ray}, P.~S., {et al.} 2012, in 2011 Fermi Symposium, arXiv:1205.3089

\bibitem[{Reynolds} {et~al.}(2007){Reynolds}, {Callanan}, {Fruchter},  {Torres}, {Beer}, \& {Gibbons}]{rcf+07}
{Reynolds}, M.~T., {Callanan}, P.~J., {Fruchter}, A.~S., {Torres}, M.~A.~P.,  {Beer}, M.~E., \& {Gibbons}, R.~A. 2007, \mnras, 379, 1117

\bibitem[{Roberts}(2011){Roberts}]{roberts11}
{Roberts}, M.~S.~E. 2011, in AIP Conf., Vol. 1357, Radio Pulsars: An  Astrophysical Key to Unlock the Secrets of the Universe, ed. {M.~Burgay,  N.~D'Amico, P.~Esposito, A.~Pellizzoni, \& A.~Possenti} (Melville, NY: AIP),  127--130, arXiv:1103.0819

\bibitem[{Romani} {et~al.}(2012){Romani}, {Filippenko}, {Silverman}, {Cenko},  {Greiner}, {Rau}, {Elliott}, \& {Pletsch}]{rfs+12}
{Romani}, R.~W., {Filippenko}, A.~V., {Silverman}, J.~M., {Cenko}, S.~B.,  {Greiner}, J., {Rau}, A., {Elliott}, J., \& {Pletsch}, H.~J. 2012, \apjl,  760, L36

\bibitem[{Shapiro} \& {Teukolsky}(1983){Shapiro} \& {Teukolsky}]{st83}
{Shapiro}, S.~L. \& {Teukolsky}, S.~A. 1983, {Black holes, white dwarfs, and  neutron stars: The physics of compact objects} (New York: Wiley-Interscience)

\bibitem[{Skrutskie} {et~al.}(2006){Skrutskie}, {Cutri}, {Stiening},  {Weinberg}, {Schneider}, {Carpenter}, {Beichman}, {Capps}, {Chester},  {Elias}, {Huchra}, {Liebert}, {Lonsdale}, {Monet}, {Price}, {Seitzer},  {Jarrett}, {Kirkpatrick}, {Gizis}, {Howard}, {Evans}, {Fowler}, {Fullmer},  {Hurt}, {Light}, {Kopan}, {Marsh}, {McCallon}, {Tam}, {Van Dyk}, \&  {Wheelock}]{2mass}
{Skrutskie}, M.~F., {et al.} 2006, \aj, 131, 1163

\bibitem[{Tauris} {et~al.}(2012){Tauris}, {Langer}, \& {Kramer}]{tlk12}
{Tauris}, T.~M., {Langer}, N., \& {Kramer}, M. 2012, \mnras, 425, 1601

\bibitem[{Tremblay} {et~al.}(2011){Tremblay}, {Bergeron}, \&  {Gianninas}]{tbg11}
{Tremblay}, P.-E., {Bergeron}, P., \& {Gianninas}, A. 2011, \apj, 730, 128

\bibitem[{Unglaub} \& {Bues}(1998){Unglaub} \& {Bues}]{ub98}
{Unglaub}, K. \& {Bues}, I. 1998, \aap, 338, 75

\bibitem[{van Haaften} {et~al.}(2012){van Haaften}, {Nelemans}, {Voss}, \&  {Jonker}]{vhnvj12}
{van Haaften}, L.~M., {Nelemans}, G., {Voss}, R., \& {Jonker}, P.~G. 2012,  \aap, 541, A22

\bibitem[{van Kerkwijk} {et~al.}(2005){van Kerkwijk}, {Bassa}, {Jacoby}, \&  {Jonker}]{vkbjj05}
{van Kerkwijk}, M.~H., {Bassa}, C.~G., {Jacoby}, B.~A., \& {Jonker}, P.~G.  2005, in ASP Conf. Ser., Vol. 328, Binary Radio Pulsars, ed. {F.~A.~Rasio \&  I.~H.~Stairs} (San Fransisco, CA: ASP), 357, arXiv:astro-ph/0405283

\bibitem[{van Kerkwijk} {et~al.}(2011){van Kerkwijk}, {Breton}, \&  {Kulkarni}]{vkbk11}
{van Kerkwijk}, M.~H., {Breton}, R.~P., \& {Kulkarni}, S.~R. 2011, \apj, 728,  95

\bibitem[{van Kerkwijk} {et~al.}(2010){van Kerkwijk}, {Rappaport}, {Breton},  {Justham}, {Podsiadlowski}, \& {Han}]{vkrb+10}
{van Kerkwijk}, M.~H., {Rappaport}, S.~A., {Breton}, R.~P., {Justham}, S.,  {Podsiadlowski}, P., \& {Han}, Z. 2010, \apj, 715, 51

\bibitem[{Vennes} {et~al.}(2011){Vennes}, {Thorstensen}, {Kawka},  {N{\'e}meth}, {Skinner}, {Pigulski}, {St{\c e}{\' s}licki},  {Ko{\l}aczkowski}, \& {{\'S}r{\'o}dka}]{vtk+11}
{Vennes}, S., {et al.} 2011, \apjl, 737, L16

\end{thebibliography}
\end{document}